\documentclass[prd,aps,showpacs,tightenlines,superscriptaddress,amsmath,amssymb,amsfonts,twocolumn]{revtex4}
\usepackage{amssymb,amsmath,amsthm,amsbsy,epsfig,color,graphicx,times}
\usepackage[ansinew]{inputenc}
\usepackage[english]{babel}
\usepackage{color}
\usepackage{braket}
\usepackage{booktabs}
\usepackage{tabularx}
\usepackage{array}
\usepackage{graphicx}
\usepackage[]{latexsym}
\usepackage{bm}

\newcommand{\be}{\begin{equation}}\newcommand{\ee}{\end{equation}}
\newcommand{\bea}{\begin{eqnarray}}\newcommand{\eea}{\end{eqnarray}}
\newcommand{\brr}{\begin{array}}\newcommand{\err}{\end{array}}
\newcommand{\bit}{\begin{itemize}}\newcommand{\eit}{\end{itemize}}
\newcommand{\ben}{\begin{enumerate}}\newcommand{\een}{\end{enumerate}}

\newcommand{\ba}{\begin{array}}
\newcommand{\ea}{\end{array}}

\def\lf{\left}

\def\non{\nonumber}

\def\ri{\right}
\def\al{\alpha}

\def\te{\theta}

\def\1{{_{1}}}\def\2{{_{2}}}

\def\noHe0{:\;\!\!\;\!\!:H_e(0):\;\!\!\;\!\!:}
\def\noHm0{:\;\!\!\;\!\!:H_\mu(0):\;\!\!\;\!\!:}
\def\nof{:\;\!\!\;\!\!:}

\def\lf{\left}

\def\non{\nonumber}

\def\ri{\right}

\def\al{\alpha}

\def\te{\theta}

\def\1{{_{1}}}\def\2{{_{2}}}
\def\nof{:\;\!\!\;\!\!:}

\def\wwQ{Q}

\begin{document}

\title{Probing dark matter and quantum field theory effects with Rydberg atoms}

\author{A. Capolupo}
\email{capolupo@sa.infn.it}
\affiliation{Dipartimento di Fisica ``E.R. Caianiello'' Universit\`{a} di Salerno, and INFN -- Gruppo Collegato di Salerno, Via Giovanni Paolo II, 132, 84084 Fisciano (SA), Italy}

\author{S. M. Giampaolo}
\email{sgiampa@irb.hr}
\affiliation{Institut Ru\dj er Bo\v{s}kovi\'c, Bijeni\v{c}ka cesta 54, 10000 Zagreb, Croatia}

\author{G. Lambiase}
\email{lambiase@sa.infn.it}
\affiliation{Dipartimento di Fisica ``E.R. Caianiello'' Universit\`{a} di Salerno, and INFN -- Gruppo Collegato di Salerno, Via Giovanni Paolo II, 132, 84084 Fisciano (SA), Italy}

\author{A. Quaranta}
\email{anquaranta@unisa.it}
\affiliation{Dipartimento di Fisica ``E.R. Caianiello'' Universit\`{a} di Salerno, and INFN -- Gruppo Collegato di Salerno, Via Giovanni Paolo II, 132, 84084 Fisciano (SA), Italy}


\begin{abstract}

 We analyze the oscillations of Rydberg atoms in the framework of quantum field theory and we reveal a non trivial vacuum energy which has the equation of state of  the dark matter.
 This energy is similar to that expected for mixed neutrinos and affects the thermal capacity of the gas.
 Therefore, deflection of the thermal capacity of Rydberg atoms could prove the condensate structure of vacuum for mixing fermions and open new scenarios in the study of the dark components of the universe.

\end{abstract}

\maketitle

\section{I. Introduction}

One of the long--lasting mysteries in current Cosmology is the nature of dark matter. First postulated to explain the rotation curve of galaxies~\cite{Trimble}, it is now believed to represent about 23\% of the total energy density of the universe~\cite{Hinshaw2013,Gupta2010}.
Despite a certain agreement on its existence, its composition is still debated.
A number of candidates for dark matter has been proposed~\cite{kam,kam1,kam2,kam3,kam4,kam5,kam6}, most focused on ``exotic particles'', yet no one of them has proven to be satisfactory enough.
However several aspects are now very clear.
Dark matter, or more precisely cold dark matter, behaves  as a pressureless perfect fluid \cite{kam}.
Accordingly with these costrains, it has been recently shown that vacuum condensates owing to the phenomenon of fermion mixing, like the flavor vacuum energy for neutrinos, satisfy the dark matter equation of state~\cite{Capolupo:2016pbg}.
This is associated with the fact that the particle mixing phenomenon, together with the Hawking or the Casimir effect etc., have a non--zero vacuum energy which cannot be removed by use of the normal ordering procedure~\cite{Hawking:1974sw,Unruh:1976db,Schwinger:1951nm,Bardeen:1957mv,Iorio:2010pv,Takahasi:1974zn,Casimir,Casimir1}.

However, a test of the contribution of the condensate of mixed fermions, like neutrinos, to the dark matter, is extremely hard to be achieved.
This is because the condensate energy is relevant only at cosmological scales and also because the fermions involved are, usually, difficult to control.
Then, it would be of great interest to individuate an analogous physical system that could be, in principle, controlled in laboratory.
Here, we identify such a system with an ensemble of independent Rydberg atoms~\cite{Lim} interacting with a laser light.
They are atoms whose outermost electron is in a large principal quantum number ($n$) excited state.
They owe their name to the fact that their energy levels closely resemble those of the hydrogen atom, up to a corrective factor called \textit{quantum defect} $\delta_l$.
Namely, for $n\gg1$ one has $ E_{n,l} = -\frac{Ry}{(n-\delta_l)^2}$, where $Ry \approx 13,6 \ eV$ is the Rydberg constant and $\delta_l$ depends both on the atomic species considered and the orbital quantum number.
Typically $\delta_l$ is significant only at low angular momentum $l \leq 4$~\cite{HJNTG,LNMG,Weber}.
For neutral alkali atoms, like Cs and Rb, the total electronic spin is $s=\frac{1}{2}$, which makes them good candidates for the simulation of a fermionic system.

The formal analogy between Rydberg atoms and neutrino mixing allows to study some features expected for mixing fermions by means of the analysis of Rydberg atoms.
Hence, we analyze the mixing of Rydberg atoms in the framework of Quantum Field Theory (QFT).
We show that the QFT formalism of oscillating Rydberg atoms is formally equivalent to that of two flavor fermion mixing and thus must produce the same condensate structure.
We prove that the energy of the physical vacuum has, indeed, the equation of state of the dark matter.
We compute the additional energy density originating from the condensate structure and its contribution to the thermal capacity of the gas.
The study of the thermal capacity of Rydberg atoms could open a new way to the research and to the understanding of the dark components of the universe.
The eventual detection of a small deviation from the expected profile of the thermal capacity of such atoms should demonstrate the existence of the condensate and should prove that a component of the dark matter is represented by quantum vacuum energy.
Moreover, we show QFT corrections to the oscillation formulas describing the transitions between ground and excited states.
Also this correction could be, in principle tested in laboratory.

The paper is structured as follows.
In Section II, we prove the formal equivalence between the QFT  of the ensemble of independent Rydberg atoms interacting with a laser and two flavor fermion mixing.
In Section III, we first compute the vacuum energy contribution due to the mixing of the Rydberg atoms and the new contribution to the thermal capacity.
Soon after we evaluate the QFT corrections to the Rabi oscillation predicted by the non--relativistic quantum mechanics.
Conclusions are reported in Section IV.


\section{QFT of Rydberg atoms}

In this section we prove the formal equivalence between the QFT formalism describing an ensemble of independent Rydberg atoms interacting with a laser light, and the QFT of two flavor fermion mixing.
This analogy can be used to study many aspects of particle mixing that are predicted theoretically, but hardly visible from an experimental point of view.

Let us start describing the system that we will consider all along the paper.
We take into account an ensemble of Rydberg atoms at a very low density, that allows us to neglect all interactions among them, and temperature low enough to permit us to discard all the finite temperature effects.
The whole set interacts with a laser light with Rabi frequency $\Omega(t)$ and detuning $\Delta(t)$, coupling two levels of the atoms that we name $g$ and $e$ with related fields $\phi_{g}$ and $\phi_{e}$, and with $E_g < E_e$.

With these assumptions the total Hamiltonian describing the system can be seen as the sum of one--particle operators each one of them describing a single atom.
The single atom Hamiltonian can be split in two contributions, one corresponding to the center of mass Hamiltonian $H_{CM}$ and the second corresponding to the outer electron $H_E$.
With these assumptions, the electronic Hamiltonian, in the rotating wave approximation, is given by~\cite{Schauss2015,Cui}
\bea
  H_{E} &  = & \ \sum_{n }\int d^{3} x (E_{n} - \Delta_{n}(t))\phi_{n}^{\dagger}(x)\phi_{n}(x)
 \\ && + \   \hbar \Omega(t) \int d^{3} x  ( \phi_{g}^{\dagger}(x) \phi_{e}(x) + \phi_{e}^{\dagger}(x) \phi_{g}(x) ) \non
\eea
where $n$ is $g$ or $e$ and $\Delta_{n}(t) = \Delta (t)\,\delta_{n e}$.
Moreover, we assume that the center of mass Hamiltonian $H_{CM}$ is that of free Dirac fields
\begin{equation}
 \!\!\!H_{CM} \! = \! \sum_{n} \!\! \int d^{3}\!x \!\left(\!\frac{i}{2}  \bar{\phi}_{n}\, \gamma^{0}\! \overleftrightarrow{\partial}_{\!0}\phi_{n} \!+ \!\frac{i}{2}  \bar{\phi}_{n} \gamma^{j}\! \overleftrightarrow{\partial}^{j}\phi_{n} \!+\! m \bar{\phi}_{n} \phi_{n} \!\right)\!\!\!\!\!\!\!\!\!\!\!\!\!\!\!\!
\end{equation}
with $m$ the mass of the atom considered.
Let us focus on the electronic part of the Hamiltonian.
By setting \mbox{$\varepsilon_{n} = E_{n} - \hbar \Delta_{n} (t) $}, we can equivalently write
\begin{equation}\label{FinalHamiltonian}
  H_E =  \int d^{3}x\, \psi^{\dagger}(x) \,\pmb{M}\, \psi (x)
\end{equation}
where $\psi^{\dagger} (x)= \begin{pmatrix}
                          \phi_g^{\dagger}(x) & \phi_e^{\dagger}(x)
                        \end{pmatrix}$
and $\pmb{M}$ is the $2\times 2$ matrix
\begin{equation}\label{MixingMatrix}
  \pmb{M} = \begin{pmatrix}
              \varepsilon_g & \hbar \Omega(t) \\
              \hbar \Omega (t) & \varepsilon_e
            \end{pmatrix}.
\end{equation}
$\pmb{M}$ is analogous to the two--flavor mass matrix which characterizes the mixing of kaons, $B^0$, $D^0$ \cite{Griffiths}, neutrinos \cite{Pontecorvo} (in two flavor mixing case), the neutron-antineutron mixing \cite{Mohapatra} and the axion-photon mixing \cite{axion}.

It is straightforward to see that the Hamiltonian in eq.~\eqref{FinalHamiltonian} can be diagonalized by a rotation
\begin{equation}\label{Primedfields}
  \begin{pmatrix}
    \phi_{g} (x) \\
    \phi_{e} (x)
  \end{pmatrix}
  =
  \begin{pmatrix}
    \cos{\theta} & \sin{\theta} \\
    - \sin{\theta} & \cos{\theta}
  \end{pmatrix}
  \begin{pmatrix}
    \phi_1 (x) \\
    \phi_2 (x)
  \end{pmatrix}
\end{equation}
where $\phi_i (x)$ (with $i = 1, 2 $) are free fields and $\theta$ is the mixing angle given by $\theta = \frac{1}{2} \arctan{(\frac{-2\hbar \Omega(t)}{\varepsilon_e - \varepsilon_g})}$.
The electron Hamiltonian is now given by
\begin{equation}\label{FreeFieldHamiltonian}
  H_E = \int d^3 x ( \varepsilon_{1} \phi^{\dagger}_{1}(x) \phi_{1} (x)  + \varepsilon_{2}\phi^{\dagger}_{2}(x) \phi_{2} (x) ) \ ,
\end{equation}
where
\begin{eqnarray}
  \varepsilon_{i} &=& \frac{\varepsilon_g + \varepsilon_e }{2} + \frac{(-1)^i}{2} \sqrt{(\varepsilon_g - \varepsilon_e)^2 + 4 \hbar^2 \Omega^2(t)}.
\end{eqnarray}
with $i=1,\,2$.
It is clear that the rotation in eq.~\eqref{Primedfields} turns the total Hamiltonian
$H = H_{CM} + H_E$ into the sum of two free Dirac Hamiltonians with masses $m_1 = m + \varepsilon_1$ and $m_2 = m + \varepsilon_2$.
Hence the atomic fields $\phi_i(x)$ admit a free field expansion
\begin{equation}\label{FreeFieldExpansion}
  \phi_{i}(x) \! = \! \sum_{r} \! \int \! \! \frac{d^3 {\bf
k}}{(2\pi)^{\frac{3}{2}}} \! \! \lf[u^{r}_{{\bf k} , i} a^{r}_{{\bf k} , i} (t) + v^{r}_{-{\bf k} , i} b^{r \dagger}_{-{\bf k} , i} (t)\ri] \! e^{i{\bf k}\cdot {\bf x}}
\end{equation}
with $u^{r}_{{\bf k},i}$, and $v^{r}_{{\bf k},i}$, solutions of Dirac equation for the fields $\phi_{i}$, with effective masses $m_i = m + \varepsilon_i$.
The annihilators and creators undergo a trivial evolution $a_{{\bf k} , i} (t) = a_{{\bf k}, i} e^{- i \omega_{{\bf k},i} t}$, with \mbox{$\omega_{{\bf k}, i} =   \sqrt{{\bf k}^2 + m_{i}^2}$}.
Introducing eq.~\eqref{FreeFieldExpansion} in the expression for the Dirac Hamiltonian, the total Hamiltonian $H = H_E + H_{CM}$ becomes
\begin{equation}\label{DiracHamiltonian}
  H  = \sum_{i,r} \int d^3 {\bf k} \, \omega_{{\bf k},i}(a^{r \dagger}_{{\bf k} , i}a_{{\bf k} , i}^r + b^{r \dagger}_{-{\bf k} , i} b_{-{\bf k} , i}^r)\,.
\end{equation}

The diagonalization of the total Hamiltonian, by means of a rotation, and the subsequent quantization of the free fields $\phi_1, \phi_2$ of eq.~\eqref{FreeFieldExpansion}, show a close analogy with the 2--flavor mixing of neutrinos~\cite{mix1,mix}.
The correspondence between the QFT of Rydberg atoms and the QFT of neutrino mixing is further established by the introduction of the mixing generator.
Indeed, the transformations in eqs.~(\ref{Primedfields}) can be recast in terms of the mixing  generator $\Xi_{\theta}(t)$ as
$ \phi_{g}(x) = \Xi^{-1}_{\bf \te}(t) \phi_{1}(x)  \Xi_{\bf \te}(t) $, and similarly for $\phi_{e}(x)$.
The mixing generator is given by
\begin{equation}
\label{generator}
  \Xi_{\bf \te}(t)\! = \!\exp\left[\theta \int d^{3}{\bf x} \left(\phi_{1}^{\dag}(x) \phi_{2}(x) - \phi_{2}^{\dag}(x) \phi_{1}(x) \right)\right]\;.
\end{equation}
This operator preserves the canonical anticommutation relations and, at finite volume, is unitary and satisfies
\mbox{$\Xi^{-1}_{\bf \te}(t)=\Xi_{\bf -\te}(t)=\Xi^{\dag}_{\bf \te}(t)$}.
The action of the generator on the creation and annihilation operators is equivalent to a Bogoliubov transformation nested into a rotation, with coefficients
$\Upsilon_{{\bf k}}  \equiv  (-1)^{r}\; u^{r\dag}_{{\bf
k},1} v^{r}_{-{\bf k},2} = -(-1)^{r}\; u^{r\dag}_{{\bf k},2}
v^{r}_{-{\bf k},1}$ and $  \Sigma_{{\bf k}}  \equiv  u^{r\dag}_{{\bf k},1}
u^{r}_{{\bf k},2} = v^{r\dag}_{-{\bf k},1} v^{r}_{-{\bf k},2}$.
Explicitly, we have
 \bea
\label{Vk}
 |\Upsilon_{{\bf
 k}}|  =  \frac{ (\omega_{k,1}+m_{1}) - (\omega_{k,2}+m_{2})}{2
 \sqrt{\omega_{k,1}\omega_{k,2}(\omega_{k,1}+m_{1})(\omega_{k,2}+m_{2})}}\, |{\bf k}|
\eea
and $\Sigma_{\bf k} = \sqrt{1-\Upsilon_{{\bf k}}^2}$.

The generator $\Xi_{\bf \te}(t)$ connects the representations associated with the ``flavor'' fields $\phi_g,\phi_e$  and the ``mass'' fields $\phi_1,\phi_2$.
In particular, the vacua in the two representations are related by means of the transformation $\ket{0(\theta,t)} \doteq \Xi_{\theta}^{-1}(t) \ket{0} $
and, in the infinite volume limit, the two representations are unitarily inequivalent~\cite{mix1,mix}.
The vacuum $\ket{0(\theta,t)}$ has the structure of a condensate of particles with definite mass, whose density is equal to
\mbox{$ \bra{0(\theta,t)} a^{\dagger}_{ {\bf k},i} a_{ {\bf k},i} \ket{0(\theta,t)}\!\! =\!\! \sin^2{\theta} | \Upsilon_{{\bf k}}|^2 \;  \forall i$}.
Namely, $ \sin^2{\theta} | \Upsilon_{{\bf k}}|^2$ represents the average number density of the atoms with momentum ${\bf k}$ in the condensate.

\section{QFT effects in Rydberg atoms}

As we have seen, the QFT describing an ensemble of independent Rydberg atoms interacting with a laser light is formally equivalent to the one associated to the two flavor neutrino mixing.
For neutrinos it has been shown that the fermion mixing leads to a non trivial contribution to the vacuum energy~\cite{Capolupo:2016pbg}, with an equation of state similar to that of the dark matter.
Then, by using the formal analogy, we show that the vacuum energy of the Rydberg atoms behaves as a dark matter component.
Such a contribution can, in principle be tested in laboratory by analyzing the specific heat of the system.
Moreover, QFT predicts a correction to the Rabi oscillations of the atoms forecast by the non--relativistic quantum mechanics.

\subsection{Dark matter--like Thermal capacity}

The expectation value of the energy momentum tensor $T^{\mu\nu}(x)$ of free
Rydberg atoms  on the vacuum for mixed fields $\ket{0(\theta,t)}$, is
$\Theta^{\mu \nu}(x)   \equiv   \bra{ 0 (\theta,t)}: T^{\mu \nu}(x): \ket{ 0 (\theta,t)}$.
Here $:...:$,  represents the normal ordering with respect to the vacuum for free fields $\ket{0}$.
The off--diagonal components of $ \Theta^{\mu \nu}(x)$ vanish, then the condensate induced by the mixing is similar to a perfect fluid.
The energy density and pressure of fermion condensates can be defined as
$ \rho = g^{00} \Theta_{00}\, $ and $ p  = - g^{jj} \Theta_{jj}$ (no summation on the index $j$ is intended), respectively.
For mixed fermions one has zero pressure, $p = 0$, and the state equation reduces to $w = p / \rho = 0$, which is that of the dark matter.
The energy density of the condensate is
\bea
\rho & = &\frac{\Delta \varepsilon \sin ^{2}\theta}{2 \pi^{2}} \int_{0}^{K} dk k^{2}
\lf( \frac{m_2 }{ \omega_{k,2}} - \frac{m_1}{ \omega_{k,1}}\ri) \,,
\eea
with $\Delta \varepsilon = \varepsilon_2 - \varepsilon_1$ and  $K$ is the cut-off on the momenta.
Explicitly, we have
\begin{equation}
\label{ener-Fer}
 \rho \! =\! \frac{\Delta \varepsilon \sin^{2}\!\theta}{2 \pi^{2}}\! \sum_i (-1)^i \!\left[\!
 K m_i \omega_{K,i}\!-\!m_i^3\ln\!\left(\!\frac{K+\omega_{K,i}}{m_i}\!\right)\!
 \right].\!\!\!\!
\end{equation}
Such expression of the energy is valid only at very low temperature of the Rydberg system since we are neglecting the finite temperature effects.
Integrating over the whole space occupied by the condensate one obtains the associated contribution to the energy $U$ and deriving with respect to the temperature $T$ one can evaluate the contribution of the condensate to the thermal capacity of the system.
Having neglected the finite temperature effects, $U$ results to be independent on the temperature.

However, a first approximation of the temperature effects can be obtained by introducing a temperature dependent cut--off.
Indeed, increasing the temperature, the average energy increases and, hence, higher momuntum states are populated.
Therefore it is reasonable that the cut--off has to increase with $T$.
In the simplest case of a linear dependence of the the cut--off $K$ on the temperature, $K(T) = a T+ K(0) $, the contribution to the specific heat at constant volume by the vacuum energy of the mixed atoms is
\bea
\label{thermal_capacity}
\!\!c_{\rho}(T) \!&\! =\! &\!  a \frac{V (m_2 - m_1) \sin^{2}\theta\, K(T)^2}{\pi^2 \sqrt{(m_{1}^2 + K(T)^2) (m_{2}^2 + K(T)^2)}}
\\ \!\!\!&\! \! & \!\times \!\left(\!m_2 \sqrt{m_{1}^2 + K(T)^2}
- \! m_1 \! \sqrt{m_{2}^2 + K(T)^2}\!\right) \non.
 \eea
where $V$ is the volume of the gas of Rydberg atoms.
We observe several unconventional aspects for this contribution to the thermal capacity.
The main one is that it is proportional to the volume of the condensate and not to the number of atoms inside.
This is due to the fact that, being associated to the vacuum energy of the condensate, it is related to the pairs of {\it virtual} Rydberg atoms that are continuosly created and annihilated out of the vacuum.
The number of such pairs is proportional to the volume occupied by the condensate and is independent on the number of {\it real} Rydberg atoms in the system.
As a consequence, being the pressure of the condensate equal to zero, from the first principle of Thermodynamics $dU=T dS$, also the entropy turns out to be related to $V$.
Indeed, this is a quantum field theoretic contribution to the entropy \cite{Umezawa} which can show counterintuitive behavior.
The contribution to the thermal capacity in eq.~(\ref{thermal_capacity}) has to be added to the ordinary terms, among which the one of a fermi gas with two Rabi coupled components.

\subsection{QFT Rabi oscillation}

The QFT of particle mixing predicts corrections to the oscillation formulae~\cite{mix1,mix}.
In virtue of the analogy we have established, similar corrections are to be expected in the oscillation formulae of Rydberg atoms interacting with a laser light.

By definition, the charge operators for ground and excited Rydberg fields, are
\bea \label{chargenorm}
&&\nof \wwQ_{ n}(t) \nof\,  \equiv  \,\int d^{3}{\bf x}\,
\nof \phi_{n}^{\dag}(x)\phi_{n}(x)  \nof \
\\\non &&=   \sum_{r}
\int d^3 {\bf k} \, \lf( \al^{r\dag}_{{\bf k}, {n}}(t)
\al^{r}_{{\bf k}, {n}}(t)\, -\, \beta^{r\dag}_{-{\bf
k}, {n}}(t) \beta^{r}_{-{\bf k}, {n}}(t)\ri)\,
\eea
where  $:\!: ... :\!:\,$ denotes normal ordering with respect to  $\ket{0 (\theta,t)}$.
The oscillation formulae describing the transitions between ground and excited states (in the Heisenberg representation) are given by
\bea\label{oscillfor1}
{\cal Q}^{\bf k}_{n \rightarrow m}(t)&=&
\langle \phi_{n}({\bf x},0)|\nof Q_{m}(t)\nof|\phi_{n}({\bf x},0) \rangle ,
\eea
Explicitly, we have ${\cal Q}^{\bf k}_{g \rightarrow g}(t)= 1-{\cal Q}^{\bf k}_{g\rightarrow e}(t) $ and
\begin{eqnarray}\label{oscillazioni1}
\label{oscillazioni2}
{\cal Q}^{\bf
k}_{g\rightarrow e}(t) \!\!&\!=\!&\!\! \sin ^{2}(2\theta )\!\left[
\! \left| \Sigma_{\mathbf{k}}\right|^{2} \! \sin
^{2}\!\left( \omega_{k}^{-}t\right) \!+\!\left| \Upsilon_{\mathbf{k
}}\right|^{2}\!\sin ^{2}\!\left( \omega_{k}^{+}t\right) \!\right] . \;\;\;\;
\end{eqnarray}
where $\omega_{k}^{\pm} = \frac{\omega _{k,2} \pm \omega _{k,1}}{2}$. Eq.~(\ref{oscillazioni1})  shows the presence of a high frequency oscillation term proportional to $|\Upsilon_{\mathbf{k}}|^{2}$ and a correction to the amplitude of the QM oscillation formulae
proportional to $| \Sigma_{\mathbf{k}}| ^{2}$.
The QFT corrections to the oscillation formulae could be, in principle, tested experimentally.

\section{ Conclusions}

We have proved that the QFT of independent Rydberg atoms interacting with a laser light is formally equivalent to that of fermion mixing.
This implies a correction to the Rabi oscillations of the former which is, at least in principle, experimentally detectable.
Furthermore, its ``flavor'' vacuum has a non trivial energy and has the state equation of the dark matter.
Exploiting such analogy, we have shown that the thermal capacity of the Rydberg atoms has a further term associated to the flavor vacuum energy.
We derived these results neglecting any finite--temperature effect.
A more realistic description requires the analysis of particle mixing at finite temperature and its contribution to the vacuum energy, which are objects of a forthcoming paper.
Our analysis indicates that a deflection of the thermal capacity from that expected for Rydberg atoms could demonstrate the existence of the condensed quantum vacuum energy. Since, in the case of neutrino mixing, the analogue vacuum energy density might represent a component of the dark sector of the universe~\cite{Capolupo:2016pbg}, our study acquires a particular importance in the context of astrophysics and cosmology.
An evidence of the existence of the vacuum condensate energy, in Rydberg atoms, would signal its presence in a variety of cases, including fermion mixing.

\section*{Acknowledgements}
A.C.  G.L. and A.Q. thank partial financial support from MIUR and INFN. A.C. and G.L. thank also the COST Action CA1511 Cosmology and Astrophysics Network for Theoretical Advances and Training Actions (CANTATA).
SMG acknowledges support from the European Regional Development Fund the Competitiveness and Cohesion Operational Programme (KK.01.1.1.06--RBI TWIN SIN) and  the Croatian Science Fund Project No. IP-2016--6--3347.
SMG also acknowledges the QuantiXLie Center of Excellence,  co--financed by the Croatian Government and European Union  (Grant KK.01.1.1.01.0004).

\end{document}